\newcommand{\CLASSINPUTtoptextmargin}{2cm}
\newcommand{\CLASSINPUTbottomtextmargin}{3cm}

\documentclass[journal, a4paper]{IEEEtran}

\usepackage{cite}

\usepackage{graphicx}
\DeclareGraphicsExtensions{.pdf}

\usepackage[cmex10]{amsmath}
\usepackage{amsfonts,amssymb,amsthm}
\usepackage{times}
\usepackage{color}

\usepackage{algorithmic}
\usepackage{algorithm}
\usepackage{microtype}

\usepackage{flushend}
\usepackage{array}
\usepackage{exscale,relsize}
\usepackage{url}

\usepackage{acronym}

\begin{document}

\setlength{\abovedisplayskip}{4pt}
\setlength{\belowdisplayskip}{4pt}
\setlength{\parskip}{0pt}

\acrodef{LSP}[LSP]{limited spectrum pool}
\acrodef{CC}[CC]{component carrier}
\acrodef{BS}[BS]{base station}
\acrodef{PDF}[PDF]{probability distribution function}
\acrodef{NE}[NE]{Nash equilibrium}
\acrodef{RAN}[RAN]{radio access network}
\acrodef{PPP}[PPP]{Poisson point process}
\acrodef{QoS}[QoS]{quality-of-service}
\acrodef{MNO}[MNO]{mobile network operator}

\title{Co-primary inter-operator spectrum
    sharing over a limited spectrum pool using repeated games}
\author{
    \IEEEauthorblockN{Bikramjit Singh\IEEEauthorrefmark{1}, Konstantinos Koufos\IEEEauthorrefmark{1}, Olav Tirkkonen\IEEEauthorrefmark{1} and Randall Berry\IEEEauthorrefmark{2}}\\
    \IEEEauthorblockA{\IEEEauthorrefmark{1}Department of Communications and Networking, Aalto University, Espoo 02150, Finland
    \\\{bikramjit.singh, konstantinos.koufos, olav.tirkkonen\}@aalto.fi}\\
    \IEEEauthorblockA{\IEEEauthorrefmark{2}Department of Electrical Engineering and Computer Science, Northwestern University, Evanston, IL 60208, USA
    \\rberry@eecs.northwestern.edu}
}

\maketitle
\thispagestyle{plain}

\begin{abstract}
We consider two small cell operators deployed in the same geographical
area, sharing spectrum resources from a common pool. A method is investigated to
coordinate the utilization of the spectrum pool without monetary
transactions and without revealing operator-specific information to
other parties. For this, we construct a protocol based
on asking and receiving \emph{spectrum usage favors} by the operators,
and keeping a book of the favors. A spectrum usage favor is exchanged
between the operators if one is asking for a permission to use some
of the resources from the pool on an exclusive basis, and the other is
willing to accept that. As a result, the proposed method does not
force an operator to take action. An operator with a high load may
take spectrum usage favors from an operator that has few users to
serve, and it is likely to return these favors in the future to show a
cooperative spirit and maintain reciprocity. We formulate the
interactions between the operators as a repeated game and determine
rules to decide whether to ask or grant a favor at each stage
game. We illustrate that under frequent network load variations, which
are expected to be prominent in small cell deployments, both operators
can attain higher user rates as compared to the case of no coordination of the resource utilization.
\end{abstract}

\section{Introduction}
\label{sec:Introduction}
In the state-of-art mobile communication systems, a network operator possesses a spectrum license that provides exclusive transmission rights for a particular range of radio frequencies. Spectrum assignment based on dedicated licenses resolves the issues related to inter-operator interference but it also results in low spectrum utilization efficiency. Inter-operator spectrum sharing is envisioned as one of the viable approaches to achieve higher operational bandwidth efficiency and meet the increasing mobile data traffic demand in a timely manner~\cite{Osseiran2014}. 

In the \ac{LSP} scenario, a limited number of operators share a common resource pool by relying on more flexible and adaptive prioritization policies than is currently possible with dedicated licenses~\cite{Lehr2008}. Cognitive radio technologies are effective measures to resolve the sharing conflicts over the \ac{LSP} under vertical spectrum sharing~\cite{Si2010Jr}, where the lessor (owner) operator has higher legacy rights over the spectrum than the lessee operator. On the other hand, the co-primary or horizontal spectrum sharing scheme conceptualizes the case where authorized operators possess equal ownership on the spectrum being adopted~\cite{Irnich2013}. However, \emph{a priori} agreements should be made on the spectrum usage with regard to the long term share of an individual operator. 

The multilateral use of shared resources in the \ac{LSP} can, for instance, be achieved with channel allocation schemes originally developed for single-operator systems. These schemes are in principle applicable to realizes inter-operator spectrum sharing, provided that the operators are willing to exchange information and cooperate honestly. Under this requirement, many spectrum sharing algorithms are available in the literature. They differ related to the domain where inter-operator interference is handled, i.e. time, frequency, and/or space.

The cooperative spectrum sharing schemes require a great deal of network information exchange among the operators, e.g., interference prices~\cite{Teng2014}, channel state information~\cite{Alsohaily2013,Wang2012}, etc., and/or employing a central entity to decide upon the resource allocation. In~\cite{Teng2014}, the operator reports the inflicted aggregate interference to the spectrum controller, and on this basis, the controller awards the spectrum pool to the impacted \acp{BS}. In~\cite{Alsohaily2013}, cooperation amongst the operators is realized by broadcasting the spectrum occupancy information, allowing small cells of competitor operators to avoid interference in accessing the spectrum pool. Similarly, operators in~\cite{Wang2012} maintain channel occupancy and spectrum reservation matrices for opportunistic access to the shared pool.

Although the achievable gains in cooperative schemes are in general high, operators may be reluctant to share proprietary  information with their competitors and may also have an incentive to mis-report this information. Finally, information exchange may incorporate excessive inter-operator signaling overhead. In this perspective, game theoretic non-cooperative schemes appear to be a more viable option to share spectrum. In these schemes, players make decisions independently; they may still cooperate with competitors but, the cooperation is entirely self-enforcing. 

In~\cite{Kamal2009, Si2010}, operators establish cooperation and play
non-zero sum games to share spectrum. However, the choice of
utility function, encompassing spectrum pricing is undesirable as it
penalizes increased spectrum usage.
In~\cite{Hailu2014}, the operators enlist their preferences of
partitioning the shared pool and the outcome is established based on a
minimum rule. This method may not work well in scenarios with
load variations, in which heavily-loaded and lightly-loaded operators will end up with the same
number of orthogonal carriers from the pool.
In~\cite{Etkin2007,Wu2009}, operators model their interactions via
repeated games. A common assumption is that they agree in advance
on the spectrum allocation, e.g., at a \ac{NE}
in~\cite{Etkin2007} or at an orthogonal allocation in~\cite{Wu2009}
and this allocation is maintained under the threat of punishment.
Auction-based sharing techniques have been discussed
in~\cite{Khaledi2013,Xu2010}, in which operators bid competitively for
spectrum access through a spectrum broker. However, operators may be
hesitant in adopting market-driven sharing schemes as they may not
want to touch their revenue model.

In this paper, we consider spectrum sharing in a setting where no
\ac{RAN} information is revealed to other operators. We assume that
operators are not willing to monetize spectrum use, keeping spectrum
sharing on the \ac{RAN} level. Unlike one-shot games, the proposed
scheme takes into account also the history of previous interactions
between the operators and entails the benefits of reciprocity. We
illustrate that a repeated game can be set up so that both operators
achieve better performance in comparison to a static spectrum
allocation. Unlike the repeated game models proposed
in~\cite{Etkin2007, Wu2009}, we do not fix the spectrum allocation but
we allow a flexible use of the \ac{LSP} based on the network load and
interference conditions. By employing the proposed scheme in a scenario
with two operators, we are able to show that under load asymmetry, both
operators can benefit as compared to a scheme where no spectrum
coordination is allowed.

The remainder of the paper is organized as follows. In Section~\ref{sec:System_Model}, we present the system model. Section~\ref{sec:Coordination} formulates the repeated game for inter-operator spectrum sharing and presents the proposed mechanism for negotiating the utilization of the spectrum pool. Section~\ref{sec:Numericals} demonstrates performance gains with the proposed scheme and finally Section~\ref{sec:Conclusions} concludes the paper with a summary and areas for further work.
  
\section{System model}
\label{sec:System_Model}
For simplicity, we concentrate on a spectrum sharing scenario with two small cell operators, Operator $A$ and Operator $B$. Each operator has one \ac{CC} for dedicated usage. The operators participate also in a \ac{LSP} and divide it into $K$ \acp{CC} of equal bandwidth. The proposed spectrum sharing scheme will be used to negotiate the utilization of the $K$ \acp{CC} in the downlink.

We consider a scenario with network load variations. At a particular time instant, the user distribution is modeled via a \ac{PPP} with a mean equal to $N_X$ users for Operator $X\!:\!X\in\{ A,B\}$. Given the network state, an operator evaluates a network utility function to describe the \ac{QoS} offered to its users. It is important to remark that operators need not employ same utility function nor to be aware of each other's utility function. For simplicity, we assume that both operators maintain a proportionally fair~\cite{Kelly1998} utility function directly constructed based on the user rates.  
\begin{equation*}
\label{eq:Utility}
U_X = \sum\limits_{n=1}^{n^{}_{X}}{\log\left (\sum\limits_{k=1}^{K+1}
    r_{n,k} \right)}  
\end{equation*}
where $n^{}_{X}$ is a particular realization of a \ac{PPP} with mean $N_X$, $r_{n,k}$ is the transmission rate of the $n$-th user of Operator $X$ on the $k$-th \ac{CC} calculated as 
\begin{equation*}
\label{eq:Rate}
r_{n,k} = w_{n,k}B \log_2\left(1+
  \frac{\gamma_{n,k}}{\gamma_{\text{eff}}}\right)   
\end{equation*}
where $w_{n,k}$ is the time scheduling weight of the $n$-th user scheduled on the $k$-th \ac{CC}, $B$ is the bandwidth of a \ac{CC}, $\gamma_{n,k}$ is the downlink user SINR and $\gamma_{\text{eff}}$ is the SINR efficiency. 

We consider downlink transmissions without power control. The downlink received signal power for the $n$-th user on the $k$-th \ac{CC} is $S_{n,k}$. Also, let us denote by $I_{n,k}$ the aggregate interference level incorporating both interference from the operator's own network, and from the other operator's interfering \acp{BS}. Then, the downlink user SINR is
\begin{equation*}
\label{eq:SINR}
\gamma_{n,k} = \frac{S_{n,k}}{I_0 + I_{n,k}}
\end{equation*}
where $I_0$ is the power per \ac{CC} of thermal noise and other interference. Note that on the dedicated \ac{CC} there is no inter-operator interference. 

The scheduling weights, $w_{n,k}$ are determined to maximize the utility $U_X$. 
\begin{equation*}
\label{eq:P1}
\begin{array}{*{20}c}
   {\mathop {{\rm{Maximize:}}}\limits_{w_{n,k} } } & {U_X.}  \\
   {{\rm{Subject}}\,{\rm{to:}}} & {\sum\limits_{n = 1}^{n^{}_X }
     {w_{n,k} }  = 1\,\forall k}  \\
   {} & {w_{n,k}  \geq 0,\,\forall \left\{ {n,k} \right\}}.
\end{array}
\end{equation*}

In order to evaluate the effect the opponent operator has on its utility, an operator may ask its users to measure the amount of aggregate interference level they receive from the opponent. This functionality requires that the users are able to separate between their own and the other operator's generated interference. Note that this kind of functionality does not require any signaling between the two operators. It is assumed that inter-operator interference measurements are ideal. 

\section{Coordination protocol}   
\label{sec:Coordination}
In small cell deployments it is expected to have changing traffic and interference profiles. Small cell deployments of different  operators sharing spectrum in the same geographical area can exploit these fluctuations and achieve mutual benefits by regulating the allocation of \acp{CC}. For instance, let us consider spectrum sharing between two operators with unbalanced traffic loads over a \ac{LSP}. A lightly-loaded operator can satisfy its \ac{QoS} with few \acp{CC} and could perhaps stop using some of the \acp{CC} from the pool. An operator that is heavily loaded at that time would not suffer from inter-operator interference on the emptied \acp{CC} and would be able to meet its \ac{QoS} too. However, there should be an incentive for the lightly-loaded operator to free up some \acp{CC}. 

Without going to monetary transactions, we propose to regulate the allocation of \acp{CC} by means of \emph{spectrum usage favors} asked and granted by the operators. In a \ac{LSP}, a spectrum usage favor refers to the following action $-$ an operator asks its competitor for permission to start using a certain number of \acp{CC} from the pool on an exclusive basis. 
While negotiating for spectrum, the operators should agree about the default utilization of the spectrum pool. In principle, any MAC protocol could be applied in the default state. We consider that both operators utilize all the \acp{CC} of the pool. A spectrum usage favor that is exchanged between the operators necessitates a departure from the default state. The time period a spectrum favor is valid is agreed between the operators. E.g., a favor can be valid in the order of seconds, reflecting the time the network states remain unchanged. After the validity of a favor expires, the utilization of the spectrum pool falls back to the default state and the operators will begin a new round of negotiations.

Since the operators will share spectrum for a long time, an operator taken favors in the past will return them in future to show a cooperative spirit and maintain the exchange of favors with its competitor. 
Monetized compensations or auction schemes are not considered here. It is a non-trivial task to design such efficient mechanisms for a limited area and a limited time, and to couple operator strategies to the income model of operators. Realizing spectrum sharing in the form of favors entails the benefits of reciprocity, circumvents monetary-based spectrum sharing and enables the operators to achieve mutual benefits without revealing RAN-specific information to competitors and/or other parties. These make the considered approach similar to peering agreements in the internet.
%
%

\subsection{One-shot game}
We first consider a one-shot game where the operators $A$ and $B$ are modeled as myopic players.
The game is strategic and non-cooperative. Each player $X$'s action or strategy, $s^{}_X$, is to either ask a favor of $k=1,\ldots,K$ CCs denoted by $(a_1,\ldots,a_k)$, grant a favor on $k=1,\ldots,K$ CCs denoted by $(g_1,\ldots,g_k)$ or do neither, denoted by $\rm n$. Let $S_X$ denote the set of such actions for player $X$.

To specify the outcome of the game, we assume that a favor is exchanged only if one player plays $g_l$ and the other plays $a_k$ with $l \geq k$, then the outcome is an exchange of $k$ \acp{CC}. Otherwise, no exchange of favors occur.\footnote{In other words, here we view the requests as being for a specific number of \acp{CC} and no fewer. Alternatively, one could study a model where the asks are for up-to that number of \acp{CC} so that if $l < k$ a trade for $l$ \acp{CC} still occurs. We leave such a model for future work.}

Depending on the outcome, operators draw rewards: (i) the reward when
a player takes a favor is equal to the utility gain when the
interference on $k$ \acp{CC} is eliminated, (ii) the reward when a
player grants a favor is equal to the utility loss when stopping to
use $k$ \acp{CC} and (iii) the reward when a player does not ask nor
grant a favor is zero. The gains and losses thus depend on the
current internal state of the player in question, and are not known to the
opponent.

In such a game, a final outcome is a \ac{NE} 
$\left(s^{}_A,s^{}_B\right) \in S_A \times S_B$,
from which no player can improve its utility by deviating unilaterally, i.e., $U_X \left( s^{}_{X} , s^{}_{-X} \right) \geq U_X \left( s^{*}_{X} , s^{}_{-X}\right)$ for every player $X = \{A,B\}$ and every alternative strategy $s^{*}_{X} \in S_X$. It is straightforward to see that the \ac{NE} of the formulated one-shot game corresponds to the situation where a player always asks for a favor on $K$ \acp{CC} to maximize its reward but never grants a favor. As a result, both operators would utilize all \acp{CC} from the pool irrespective of the network load and interference profiles.

The game formulated here differs from the power control game in
\cite{Etkin2007} in that here the actions are requests of other player
powers, and power control is binary, whereas in \cite{Etkin2007}, the
actions are power allocation profiles across frequency. Also, no
information of other operator state is assumed here. The \acp{NE} of the
strategic one-shot games discussed here and in \cite{Etkin2007},
however, are similar; full usage of the available spectrum is the only
option for a rational player.

\subsection{Repeated game}
Since the operators will share spectrum for a long time, the one-shot
game described above will be played repeatedly. In a repeated game,
the action of a player at a stage game depends not only on the current
rewards but also in the sequence of previous rewards see, e.g.,~\cite{Osborne}. 
Repeated games, such as prisoner's dilemma have been well-studied and admit a rich set of equilibrium profiles~\cite{Osborne}  including various punishment strategies. The setting here is even more challenging as the game we are interested in is a {\it stochastic game} meaning that each player's pay-off depends on a random parameter, namely the configuration of the users at that time, and moreover the player's have {\it imperfect information} since they only observe their own user configuration. We thus have a {\it Bayesian game}. Given this, we focus on a simplified set of stationary threshold policies and characterize an equilibrium among these policies.

Negotiations of favors on a single \ac{CC} were considered in~\cite{Bikram2014}. Here, we extend that to a situation with multiple \acp{CC}.
At each stage of the game an operator can compute its utility gain and utility loss by asking and granting favors on $k$ \acp{CC} for $k=1,\ldots, K$. The \ac{PDF} of the utility gains when Operator $X$ gets a favor on $k$ \acp{CC} is denoted by $f_{X,G_k}$ and similarly, the \ac{PDF} of utility losses by $f_{X,L_k}$. In our system setup, the randomness is only due to the Poisson distribution of the operator's own users (and the user's corresponding channel gains). Therefore the \acp{PDF} depend only on the network state of the operator's own network. However, note that if power control and inter-cell interference coordination is employed, these distributions may depend on the state of the opponent operator's network too. 

At each stage game, the Operator $X$ first checks whether to ask for a
favor on $K$ \acp{CC} by comparing its immediate utility gain with a
threshold $\theta_{X,K}$. If the utility gain is less than the
threshold $\theta_{X,K}$, the operator can consider asking for a favor
on $(K-1)$ \acp{CC} instead, and so forth. 
We assume that an operator always asks for the largest number of \acp{CC} for which its utility gain exceeds the corresponding threshold. 
As a result, the
probability that Operator $X$ asks for a favor on $k$ \acp{CC} is
equal to the probability that the utility gain from taking a favor on
$\{j:j=(k+1),\ldots, K\}$ \acp{CC} is less than the corresponding
thresholds $\theta_{X,j}$ and also, the utility gain from taking a
favor on $k$ \acp{CC} is higher than the threshold $\theta_{X,k}$
\begin{equation}
\label{eq:asks}
{P}^{\rm ask}_{X,k} = \prod_{j = k+1}^K \int\nolimits_0^{\theta_{X,j}} \!\!\!\! {f_{X,G_j}\left( g_j \right)dg_j} \int\nolimits_{\theta_{X,k}}^\infty \!\!\!\! {f_{X,G_k}\left( g_k \right)dg_k} 
\end{equation}
where, to simplify the analysis, it has been assumed that the distributions of utility gains from taking favors on different number of \acp{CC} are independent. 

Similarly, the Operator $X$ grants a favor on $k$ \acp{CC} upon being asked, if its immediate utility loss is smaller than a threshold $\lambda_{X,k}$ and it has not already requested a favor. Taking into account the fact that an operator cannot ask and grant a favor at the same stage game, the probability to grant a favor on $k$ \acp{CC} is 
\begin{equation}
\label{eq:grants}
{P}^{\rm grant}_{X,k} = \prod_{j = 1}^K \int\nolimits_0^{\theta_{X,j}} \!\!\!\! {f_{X,G_j}\left(g_j\right)dg_j}\int_0^{\lambda_{X,k}} \!\!\!\!{f_{X,L_k}\left( {l_k} \right)dl_k}
\end{equation}
where it has been assumed that the distributions of utility gains and utility losses are independent. 

We assume that the networks of the operators are similar,
and in symmetric relationship with each other, and do not assume a
discount of favors. To get preliminary understanding on steady
state behavior in such a setting, inspired by~\cite{Cheng2004} we thus assume that
averaged over long times, operators give and take the same amount of equally valuable favors. Hence, favors would become a rudimentary RAN-level spectrum sharing currency. Thus we have
\begin{equation}
\displaystyle\sum_{k=1}^{K} k  {P}^{\rm ask}_{A,k} {P}^{\rm grant}_{B,k} = \displaystyle\sum_{k=1}^{K} k  {P}^{\rm ask}_{B,k}  {P}^{\rm grant}_{A,k}
\label{eq:constraint}
\end{equation}
where the left-hand side describes the average number of \acp{CC} that Operator $A$ gets a favor on, and the right-hand side the same quantity for Operator $B$. 

An operator can monitor the probabilities of asking and granting of the opponent and set its own decision thresholds for satisfying the constraint~\eqref{eq:constraint}. However, there may be multiple combinations of thresholds fulfilling the constraint. We propose to  identify the thresholds maximizing an excess utility ${\mathop U\limits^\sim}$ calculated over the \ac{NE} of the one-shot game (i.e. both operators utilize simultaneously all the $K$ \acp{CC}). The excess utility for an operator reflects its expected gain from taking a favor penalized by its expected loss from granting a favor. In order to avoid unnecessary complexity in the notation we show how to set the decision thresholds for Operator $A$. Similarly, the decision thresholds for Operator $B$ can be computed.  

The excess utility for Operator $A$ is 
\begin{equation}
\label{eq:Ua}
{{\mathop U\limits^\sim}_A}  =\displaystyle \sum_{k=1}^{K} \left( \Gamma_{A,k}  - \Lambda_{A,k}  \right)
\end{equation}
where $\Gamma_{A,k}$ and $\Lambda_{A,k}$ are the average gain and loss in utility on $k$ \acp{CC} for Operator $A$ such that,
\begin{eqnarray}
\Gamma_{A,k}  \!\!\!\!&=& \!\!\!\! {P}^{{\rm grant}}_{B,k}  \!\! \displaystyle\prod_{j = k+1}^{K}  \!\! \int\nolimits_0^{\theta_{{A,j}}}\!\!\!\!\!\! {f_{A,G_j}\left( g_{j} \right)dg_{j}}  \!\! \displaystyle \int_{\theta_{{A,k}}}^\infty\!\!\!\!\!\!{g_{k} f_{A,G_k}(g_{k}) dg_{k}}, \nonumber \\
\Lambda_{A,k}  \!\!\!\!  &=&  \!\!\!\! {P}^{\rm ask}_{B,k}  \!\! \displaystyle \int_0^{{\lambda_{A,k}}}\!\!\!\!\!\! {l_{k}f_{A,L_k}\left( {l_{k}} \right)dl_{k}}\! \displaystyle \prod_{j = 1}^{K} \int\nolimits_0^{\theta_{A,j}}\!\!\!\!\!\!{f_{A,G_j}\left( g_j\right)dg_j}. \nonumber
\end{eqnarray}

The optimization problem for identifying the decision thresholds is 
\begin{equation}
\label{eq:P0}
\begin{array}{*{20}c}
   {\mathop {{\rm{Maximize:}}}\limits_{\theta_{A,k},\lambda_{A,k} \; \forall k}} & {{{\mathop U\limits^\sim}_A}},\\
   {{\rm{Subject}}\,{\rm{to:}}}   & {\rm{Eq.}}~\eqref{eq:constraint}. 
\end{array}
\end{equation}

In order to solve this optimization problem, we construct the Lagrangian function and solve the system of first-order conditions. 
\begin{equation*}
\mathcal{L}_A = {{\mathop U\limits^\sim}_A} - \mu_A \sum_{k=1}^K k \left( {P}^{\rm ask}_{A,k} {P}^{\rm grant}_{B,k} - {P}^{\rm ask}_{B,k} {P}^{{\rm grant}}_{A,k} \right)
\end{equation*}
where $\mu_A$ is the Lagrange multiplier. 

Starting with the partial derivative of the Lagrangian in terms of $\lambda_{A,1}$ and setting it equal to zero allows computing the value of the Lagrange multiplier $\mu_A=\lambda_{A,1}$. Setting the partial derivative of the Lagrangian with respect to $\lambda_{A,k}:k>1$ equal to zero, and substituting the value of the Lagrange multiplier into the resulting equation gives 
\begin{equation}
\lambda_{A,k}=k\lambda_{A,1},\, k\!>\!1.
\label{eq:dervln}
\end{equation}

Next, starting from $\partial\mathcal{L}_A / \partial\theta_{A,1}=0$ one can determine the threshold $\theta_{A,1}$ as a function of the thresholds $\lambda_{A,k}$ 
\begin{equation}
\label{eq:dervg1}
{P}^{{\rm grant}}_{B,1}\!\!\left(\theta_{A,1}\!-\!\lambda_{A,1}\right)\!=\!\!\displaystyle\sum\limits_{j=1}^K \!{P}^{\rm ask}_{B,j} \!\!\displaystyle\int_0^{\lambda_{A,j}}\!\!\!\!\!\!\!\left(\lambda_{A,j}\!-l_{j}\right)\!f_{A,L_j}\!\left(l_j\right)\! dl_{j}.\!
\end{equation}

Finally, setting $\partial\mathcal{L}_A / \partial\theta_{A,k}\!=\!0,k\!>\!1$ and using the solution for $\theta_{A,k-1}$, we end up with 
\begin{eqnarray}
\label{eq:dervgn}
\!\!\!\!\theta_{A,k} \! \!\!\!\! &=&\!\!\!\! \!\lambda_{A,k} \nonumber\\
& &\!\!\!\! \!+\frac{{P}^{{\rm grant}}_{B,k-1}} {{P}^{{\rm grant}}_{B,k}} \!\! \left(\! \displaystyle \int_{\theta_{A,k-1}}^\infty\!\!\!\!\!\!\!\!\!\!\!\!\!\!\!\left(g_{k-\!1}\!-\!\lambda_{A,k-\!1}\right) \!f_{A,G_{k-1}}\!\! \left(g_{k-1}\!\right)dg_{k-\!1} \right.\!\!\!\!\!\!\!\!\!  \\
& &\;\;\;\;\;\;\;\;\;\;\;\;\left.+\!\left(\theta_{A,k-\! 1}\!\!-\!\!\lambda_{A,k-\! 1}\right) \!\!\!\displaystyle \int_0^{\theta_{A,k-1}}\!\!\!\!\!\!\!\!\!\!\!\!\!\!\!\!\!f_{A,G_{k-\!1}}\!\! \left(g_{k-\!1}\!\right)\! dg_{k-\!1}\! \!\right)\!\!.\nonumber
\end{eqnarray}

The thresholds $\lambda_{A,k},\theta_{A,k} \forall k$ that may maximize the Lagrangian must jointly satisfy equations~\eqref{eq:dervln}$-$\eqref{eq:dervgn} and also the constraint~\eqref{eq:constraint}. Note that the above system of equations does not accept a closed-form solution but it is straightforward to solve numerically. Also, from equations (\ref{eq:dervg1},\ref{eq:dervgn}) one can deduce that $\theta_{A,k}\!>\!\lambda_{A,k}\, \forall k$. In the Appendix, we show that the solution satisfying the first-order conditions and the constraint satisfies ${\mathop U\limits^\sim}_A \!>\!� 0$. Therefore, the proposed method can achieve better performance in comparison to the \ac{NE} of the one-shot game. Finally, besides the calculation of the Lagrangian at the stationary point, we also compute it at the borders. The thresholds, either interior or border, maximizing the Lagrangian are selected. 

\section{Numerical Examples}
\label{sec:Numericals}
\begin{figure}[t]
  \centering
  \includegraphics[trim = 0mm 0.5mm 0mm 1.5mm,clip,width=0.34\textwidth]{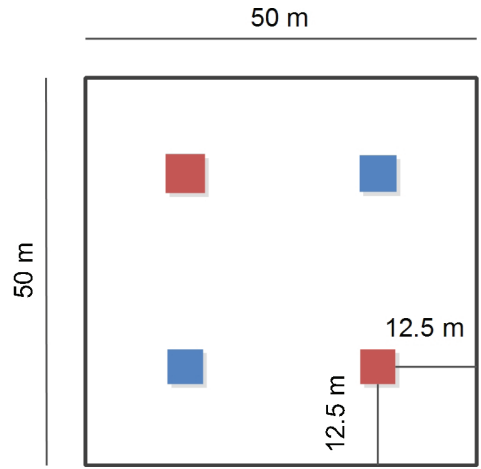}
  \caption{Indoor inter-operator deployment scenario. 
	Different colors represent \acp{BS} of different operators.} 
  \label{fig:Building}
\end{figure}

In order to assess the performance of the proposed coordination protocol, we consider an indoor deployment scenario in a hall of a single-story building. The hall is a square with a side of $50$ m. The \acp{BS} are partitioned into two groups as illustrated in Fig.~\ref{fig:Building} modeling a spectrum sharing scenario with two operators. The service areas of the operators fully overlap. A user is connected to the \ac{BS} of its home network with the highest received signal level at its location. 
We consider a power law model for distance-based propagation pathloss with attenuation constant $10^{-4}$ and pathloss exponent $3.7$. The available power budget on a \ac{CC} is $20$ dBm, the thermal noise power is $-174$ dBm/Hz and the noise figure is $10$ dB. The SINR efficiency is $\gamma_{eff}\!=\!2$. The bandwidth of a \ac{CC} is $B\!=\!20$ MHz. Initially, we consider that the \ac{LSP} consists of two \acp{CC}.

\begin{figure}[t]
  \centering
  \includegraphics[trim = 0mm 2.5mm 0mm 7.5mm,clip,width=0.44\textwidth]{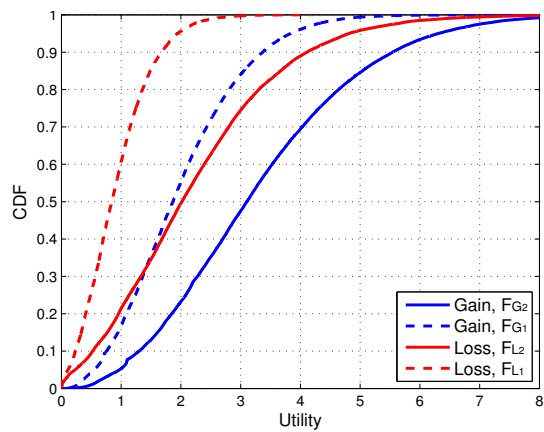}
  \caption{Distribution of utility gains and utility losses for Operator $X$ from spectrum usage favors over one and two \acp{CC} at the end of the initialization.} 
  \label{fig:GainLossDistr}
\end{figure}
\begin{figure}[t]
  \centering
  \includegraphics[trim = 0mm 2.5mm 0mm 7mm,clip,width=0.44\textwidth]{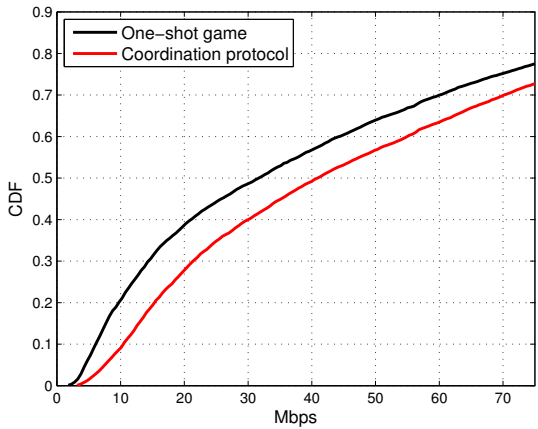}
  \caption{Rate distribution for the users of Operator $A$ obtained by the proposed scheme and by static spectrum allocation
scheme where the operators always utilize the pool consisting of two \acp{CC}. Unequal mean network loads for the two operators.} 
  \label{fig:OprA_Asy}
\end{figure}

First, we consider an initialization phase of $200\, 000$ simulation snapshots (or equivalently $200\,000$ stage games). At each stage game, the user locations are independently generated according to the \ac{PPP} and the operators calculate and keep track of their utility gains and utility losses from taking and granting favors over one and two \acp{CC}. We simulate many different network loads so that the distributions of utility gains and utility losses at the end of the initialization can be seen as the steady state distributions over all possible network states. The different network loads are generated by varying the mean of the \ac{PPP} used to model the locations of the users. In Fig.~\ref{fig:GainLossDistr} we depict the distributions of utility gains and losses for an operator at the end of the initialization.

Next, we evaluate the performance in terms of the user rate distribution over a finite time horizon of $10\,000$ stage games following the initialization phase. Initially, the values of the thresholds are set arbitrarily equal to $\theta_{X,2}\!=\!3$, $\theta_{X,1}\!=\!1.5$, $\lambda_{X,2}\!=\!2$ and $\lambda_{X,1}\!=\!1$ for both operators. Every $100$ stage games, the operators'  probabilities for asking and granting favors are recomputed considering all stage games. Then, the decision thresholds are updated by solving the optimization problem~\eqref{eq:P0} and so forth. Given the allocation of \acp{CC} at each stage of the game, the operators compute and keep track of the user rates. Recall that granted favors are valid only for a particular stage game. At the end of each stage game, the \ac{CC} allocation returns to the default state i.e. both operators utilize all the \acp{CC} of the \ac{LSP}. The performance of the proposed scheme is assessed in comparison with the \ac{NE} of the one shot game, which is a static spectrum allocation scheme where both operators utilize all the \acp{CC} of the \ac{LSP}.
\begin{figure}[t]
  \centering
  \includegraphics[trim = 0mm 2.5mm 0mm 7mm,clip,width=0.44\textwidth]{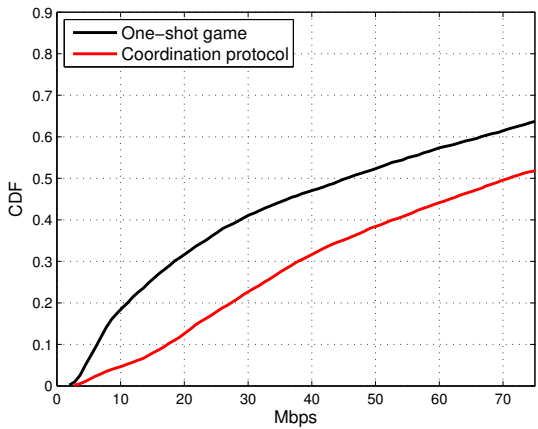}
  \caption{Rate distribution for the users of Operator $A$ obtained by the proposed scheme and by static spectrum allocation.  Four \acp{CC}. Unequal mean network loads for the operators.} 
\label{fig:OprA_Asy_4SCC}
\end{figure}

First, we consider a scenario with network load asymmetry between the operators. The mean number of users for the first $5\,000$ stage games are $N_A\!=\!8$ and $N_B\!=\!2$. In the second half of the simulation, the mean values are reversed. In  Fig.~\ref{fig:OprA_Asy}, the rate distribution for the users of Operator $A$ is depicted over the full course of the simulation. In the first $5\,000$ stage games, Operator $B$ can mostly cope with fewer or no \acp{CC} due to the lower load and it grants more favors than the Operator $A$. In the second half of the simulation, Operator $A$ returns the favors. Overall, Operator $A$  offers better \ac{QoS} in comparison with the \ac{QoS} attained without any coordination e.g. it improves its mean user rate by approximately $28$ \%. The user rate distribution curves for Operator $B$ follow the same trend and are not depicted. Fig.~\ref{fig:OprA_Asy_4SCC} depicts the rate distribution curves for the users of Operator $A$ when the \ac{LSP} consists of four \acp{CC}. One can see that the mean user rate increases by approximately $50$ \% while, the user rate at the $10$ \% of the distribution increases by more than $100$ \%. 

\begin{figure}[t]
  \centering
  \includegraphics[trim = 0mm 2.5mm 0mm 7mm,clip,width=0.44\textwidth]{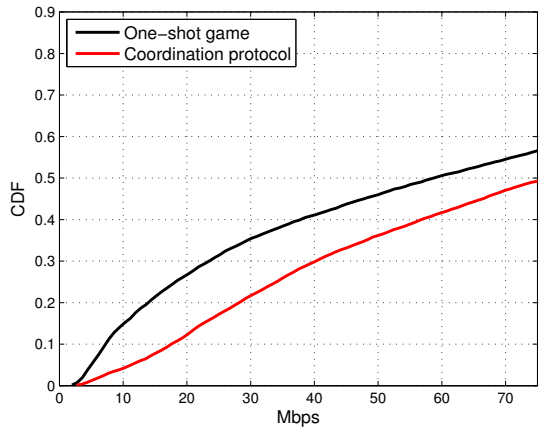}
  \caption{Rate distribution for the users of Operator $A$ obtained by the proposed scheme and also by static spectrum allocation. Four \acp{CC}. Equal mean network loads for the operators.} 
  \label{fig:OprA_Sym}
\end{figure}

Finally, we show that gains due to coordination can be achieved even in cases with equal mean network loads for the operators. In that case, the proposed protocol takes advantage of the instantaneous network load variations. In Fig.~\ref{fig:OprA_Sym} we see that Operator $A$ improves its mean user rate by approximately $25$ \% for mean number of users $N_A \!=\! N_B\!=\! 5$ and four \acp{CC} in the \ac{LSP}. 

\section{Conclusions}
\label{sec:Conclusions}
In this paper, we considered co-primary spectrum sharing between two small cell operators deployed in the same geographical area. We considered a scenario where the operators have equal access rights on a spectrum pool and we proposed a protocol for coordinating the utilization of component carriers from the pool. According to it, an operator may ask for spectrum usage favors from its competitor. A spectrum usage favor means that the competitor would stop using some component carriers from the pool. An operator that has few users to serve could perhaps cope with less component carriers and grant the favor. Operators that have taken favors in the past are likely to return these favors in future and reciprocity is maintained. We formulated the interaction between the operators as a strategic, non-cooperative repeated game. Since it is hard to analyze the proposed game and find its Nash equilibrium, we resort to a heuristic strategy that uses a threshold-based test to decide whether to ask or grant a favor at each stage game. The decision thresholds depend on the current network realization and also in the history of previous interactions with the competitor. We proved that the proposed strategy is strictly better as compared to the case without coordination between the operators. We illustrated that in an indoor deployment scenario, two operators are both able to offer higher user rates as compared to the case with no coordination, without revealing any operator-specific information to each other. Our results show that a rational operator, knowing that the opponent is rational and has a network with similar characteristics, has incentive to be cooperative. 
In future works, operators with non-similar load and network characteristics will be addressed, as well as models where the statistics of the underlying Poisson process changes to reflect e.g. variations in load due to time-of-day.

\section{Appendix}
Using the fact that the decision thresholds over the distribution of utility losses are related as $\lambda_{A,k} = k \lambda_{A,1} \forall k: k>1$, one can write write equation~\eqref{eq:dervg1} as 
\begin{eqnarray}
\label{eq:dervg1_rewrite}
\!\! \displaystyle\sum\limits_{k=1}^K \! {P}^{\rm ask}_{B,k}\!\displaystyle\int_0^{\lambda_{A,k}}\! \! \!\!\!\!\!\!\! l_{k}f_{A,L_k}\left(l_k\right)dl_k\!\!\!\!  \! &=& \!\!\! \! \! \lambda_{A,1}\! \displaystyle\sum\limits_{k=1}^K \! k {P}^{\rm ask}_{B,k} \!\displaystyle \int_0^{\lambda_{A,k}}\! \!\!\!\!\!\!\!\!f_{A,L_k}\left(l_k\right)dl_k \nonumber\\
& &\!\!\! \! \!  - \;{P}^{{\rm grant}}_{B,1}\left(\theta_{A,1}-\lambda_{A,1}\right).
\end{eqnarray}

Using equation~\eqref{eq:dervg1_rewrite} into equation~\eqref{eq:Ua}, the excess utility can be read as
\begin{eqnarray}
\label{eq:Ua_rewrite}
{{\mathop U\limits^\sim}_A}\!\!\!\! \!\!&=&\!\!\!\!\!\! \displaystyle \sum_{k=1}^{K}\! \left( \!{P}^{\rm grant}_{B,k}\!\!\! \displaystyle\prod_{j = k+1}^{K}\! \int\nolimits_0^{\theta_{A,j}}\!\!\!\!\!\!\!\!\! {f_{A,G_j}\left( g_j\right)dg_j}\!\displaystyle \int_{\theta_{A,k}}^\infty\!\!\!\!\!\!\!\! {g_kf_{A,G_k}\left( g_k\right)dg_k}\!\right) \nonumber \\ 
& & \!\!\!\! \!\!+\; {P}^{\rm grant}_{B,1}\!\left(\theta_{A,1}-\lambda_{A,1}\right) \displaystyle \prod_{j = 1}^{K} \int\nolimits_0^{\theta_{A,j}} \!\!\!\!\!\!\!\! {f_{A,G_j}\left( g_j \right)dg_j} \\
& & \!\!\!\!\!\!- \;\lambda_{A,1}\!\displaystyle\sum\limits_{k=1}^K k {P}^{\rm ask}_{B,k}\! \displaystyle\int_0^{\lambda_{A,k}}\!\!\!\!\!\!\!\! \!f_{A,L_k}\left(l_k\right)dl_k\displaystyle \prod_{j = 1}^{K} \int\nolimits_0^{\theta_{A,j}}\!\!\!\!\!\!\!\!\! {f_{A,G_j}\left( g_j \right)dg_j}. \nonumber
\end{eqnarray}

According to the definition of the probabilities of granting a favor from equation~\eqref{eq:grants}, we note that the last term in equation~\eqref{eq:Ua_rewrite} is equal to the right-hand side of the constraint in equation~\eqref{eq:constraint} scaled by $\lambda_{A,1}$. After replacing the last term of equation~\eqref{eq:Ua_rewrite} by the left-hand side of the constraint in equation~\eqref{eq:constraint}, we end up with 
\begin{eqnarray}
\label{eq:Ua_rewrite2}
\!{{\mathop U\limits^\sim}_A}\!\!\!\!\!\!&=& \!\!\!\!\!\! \displaystyle \sum_{k=1}^{K} \!\! \left( \!\!{P}^{{\rm grant}}_{B,k} \!\!\displaystyle\prod_{j = k+1}^{K} \!\int\nolimits_0^{\theta_{A,j}}\!\!\!\!\!\!\!{f_{A,G_j}\!\!\left( g_j \right)dg_j}\!\!\!\displaystyle \int_{\theta_{A,k}}^\infty\!\!\!\!\!\!\!{g_k f_{A,G_k} \!\left( g_k\right) \! dg_k}\!\!\right) \!\! \! \nonumber \\ 
& & \!\!\!\!\!\! +\; {P}^{\rm grant}_{B,1}\!\left(\theta_{A,1}-\lambda_{A,1}\right) \!\displaystyle \prod_{j = 1}^{K} \! \int\nolimits_0^{\theta_{A,j}}\!\!\!\! \!\!\! {f_{A,G_j}\!\left( g_j \right)dg_j}\\
& & \!\!\!\!\!\! - \;\lambda_{A,1}\!\!\displaystyle\sum_{k=1}^{K} k \; {P}^{\rm ask}_{A,k} \; {P}^{\rm grant}_{B,k}. \nonumber
\end{eqnarray}

Using the probabilities of ask a favor from equation~\eqref{eq:asks} into the last term of equation~\eqref{eq:Ua_rewrite2} and replacing back the decision thresholds, $\lambda_{A,k}=k\lambda_{A,1}$, 
the first and the last terms of equation~\eqref{eq:Ua_rewrite2} can be factorized together resulting to
\begin{eqnarray*}
{{\mathop U\limits^\sim}_A} \!\!\!\!\!  \! &=& \!\!\!\!\!\! \displaystyle \sum_{k=1}^{K} \! \!\left( \!\!{P}^{{\rm grant}}_{B,k} \!\!\!\!\! \displaystyle\prod_{j = k+1}^{K} \! \int\nolimits_0^{\theta_{A,j}}\! \!\!\!\!\!\!\!\!\!\!\!{f_{A,G_j}\! \left( g_j \right) \! dg_j}\!\!\! \displaystyle \int_{\theta_{A,k}}^\infty\!\!\!\!\!\!\!\!\!\!{\left(g_k \! - \! \lambda_{A,k} \right) \! f_{A,G_k} \!\left(  g_k \! \right)\!dg_k} \!\!\right) \!  \\ 
& & \!\!\!\!\!\! +\; {P}^{\rm grant}_{B,1}\left(\theta_{A,1}-\lambda_{A,1}\right) \displaystyle \prod_{j = 1}^{K} \int\nolimits_0^{\theta_{A,j}} \!\!\!\!\!\!\!\!\! {f_{A,G_j}\left( g_j \right)dg_j}.
\end{eqnarray*}
which is always positive since $\theta_{A,k}>\lambda_{A,k}\; \forall k$.

\section*{Acknowledgment}
This work has been performed in the framework of the FP7 project ICT
317669 METIS, which is partly funded by the European Union. Also, 
this work was supported in part by the Academy of Finland funded
project SMACIW under Grant no. 265040.

\small{}


\begin{thebibliography}{1}

\bibitem{Osseiran2014} A.~Osseiran \emph{et al.}, ``Scenarios for 5{G} mobile and wireless communications: {T}he vision of the {METIS} project," \emph{IEEE Commun. Mag.}, vol.~52, no.~5, pp.~26-35, May 2014.

\bibitem{Lehr2008} W. Lehr and N. Jesuale, ``Spectrum pooling for next generation public safety radio systems," in \emph{Proc. IEEE DySPAN},  pp.~1-23,~Oct. 2008.

\bibitem{Si2010Jr} P. Si, H. Ji, F.~Yu and V.~Leung, ``Optimal cooperative internetwork spectrum sharing for cognitive radio systems with spectrum pooling," \emph{IEEE Trans. Veh. Technol.}, vol.~59, no.~4, pp.~1760-1768, May 2010.

\bibitem{Irnich2013} T.~Irnich, J.~Kronander, Y.~Sel\'{e}n and G. Li, ``Spectrum sharing scenarios and resulting technical requirements for 5{G} systems," in \emph{Proc. IEEE PIMRC}, pp.~127-132, Sep.~2013.



\bibitem{Teng2014} Y.~Teng, Y.~Wang and K.~Horneman, ``Co-primary spectrum sharing for denser networks in local area," in \emph{Proc. CROWNCOM}, pp.~120-124,~Jun. 2014.

\bibitem{Alsohaily2013} A.~Alsohaily and E.~S.~Sousa, ``Spectrum sharing {LTE}-advanced small cell systems," in \emph{Proc. WPMC}, pp.~1-5,~Jun. 2013.

\bibitem{Wang2012} P.~Wang, B.~Wang, W.~Wang, Y.~Zhang and C.~Wang, ``Based multi-operator Shared Network Opportunistic Spectrum Sharing," in \emph{Proc. IEEE CCIS}, pp.~864-868, Oct. 2012.

\bibitem{Kamal2009} H.~Kamal, M.~Coupechoux and P.~Godlewski, ``Inter-operator spectrum sharing for cellular networks using game theory," in \emph{Proc. IEEE PIMRC}, pp.~425-429,~Sep. 2009.

\bibitem{Si2010} P. Si, E. Sun, R. Yang and Y. Zhang, ``Cooperative and distributed spectrum sharing in dynamic spectrum pooling networks,"  in \emph{Proc. WOCC}, pp.~1-5,~May 2010.

\bibitem{Hailu2014} S.~Hailu, A.~A.~Dowhuszko, O.~Tirkkonen and L. Wei, ``One-shot games for spectrum sharing among co-located radio access networks," in \emph{Proc. IEEE ICCS}, pp.61-66,~Nov. 2014.

\bibitem{Etkin2007} R.~Etkin, A.~Parekh and D.~Tse, ``Spectrum sharing for unlicensed bands," \emph{IEEE J. Sel. Areas Commun.}, vol.~25, no.~3, pp.~517-528, Apr. 2007.

\bibitem{Wu2009} Y.~Wu, B.~Wang, K.~Liu and T.~Clancy, ``Repeated open spectrum sharing game with cheat-proof strategies," \emph{IEEE Trans. Wireless Commun.}, vol.~8, no.~4, pp.~1922-1933, Nov.~2009.


\bibitem{Khaledi2013} M.~Khaledi, A.~A.~Abouzeid, ``A reserve price auction for spectrum sharing with heterogeneous channels," in \emph{Proc. ICCCN}, pp.~1-7, Jul. 2013.

\bibitem{Xu2010} W.~Xu and J.~Wang, ``Double auction based spectrum sharing for wireless operators," in \emph{Proc. IEEE PIMRC}, pp.~2650-2654,~Sep. 2010.

\bibitem{Kelly1998} F.~P.~Kelly, A.~K.~Maulloo, and D.~K.~H.~Tan, ``Rate control for communication networks: shadow prices, proportional fairness and stability," \emph{J. Operational Research Soc.}, v.~49, no.~3, pp.~237-252, 1998.


\bibitem{Osborne} M.~J.~Osborne, \emph{An Introduction to Game Theory}, Oxford: Oxford University Press, 2003.

\bibitem{Bikram2014} B.~Singh, K.~Koufos and O.~Tirkkonen, ``Co-primary inter-operator spectrum sharing using repeated games," in \emph{Proc. IEEE ICCS}, pp.~67-71, Nov. 2014.

\bibitem{Cheng2004} S.~F.~Cheng, D.~Reeves, Y.~Vorobeychik and M.~Wellman, ``Notes on equilibria in symmetric games," in~\emph{Proc. Workshop on Game Theoretic and Decision Theoretic Agents}, pp~71-78, Aug. 2004. 

%
%
%
%
%
%
%
%
%
%
%


\end{thebibliography}
\end{document}